\documentstyle[array,multicol,aps,rmp,psfig]{revtex}
\tighten
\begin{document}

\title{\Large \bf 
Instantons in the working memory: implications for schizophrenia.
}
\author{Alexei A. Koulakov}
\address{Sloan Center for Theoretical Neurobiology, The Salk Institute, La Jolla, CA 92037}

\date{today}
\draft
\tighten
\maketitle

\begin{abstract}
The influence of the synaptic channel properties 
on the stability of delayed activity maintained 
by recurrent neural network is studied. The duration
of excitatory post-synaptic current (EPSC) is shown
to be essential for the global stability of the 
delayed response. NMDA receptor channel is a much more reliable
mediator of the reverberating activity than AMPA receptor, due to a longer EPSC.
This allows to interpret the deterioration of working memory 
observed in the NMDA channel blockade experiments.
The key mechanism leading to the decay of the 
delayed activity originates in the unreliability
of the synaptic transmission. The optimum fluctuation
of the synaptic conductances leading to the decay is identified.
The decay time is calculated analytically and the result
is confirmed computationally. 
\end{abstract}

\begin{multicols}{2}


\section{INTRODUCTION}

Long-term memory is thought to be stored in biochemical modulation
of the inter-neuron synaptic connections. As each neuron forms about $10^4$
synapses such a mechanism of memory seems to be very effective, allowing
potentially the storage of $>10^4$ bits per neuron. 
However, although the synaptic modifications persist for a long time,
it may take as long as minutes to form them.
Since the external environment operates at much shorter time scales,
the synaptic plasticity is virtually useless for the short-term needs of survival.
Such needs are satisfied by the working memory (WM), which is stored in the state of neuronal activity, 
rather than in modification of synaptic conductances 
(Miyashita, 1988; Miyashita and Chang, 1988; Sakai and Miyashita, 1991; Funahashi {\it et al.,} 1989;
Goldman-Rakic {\it et al.,} 1990; Fuster, 1995).
 
WM is believed to be formed by recurrent neural circuits (Wilson and Cowan, 1972;
Amit and Tsodyks, 1991a, b; Goldman-Rakic, 1995).
A recurrent neural network can have two stable states (attractors),
characterized by low and high firing frequencies. 
External inputs produce transitions between these states.
After transition the network maintain high or low firing frequency during the delay period.
Such a bipositional ``switch'' can therefore store one bit of information. 

During the last few years it has become clear that NMDA receptor
is critically involved in the mechanism of WM. 
It is evident from the impairments of the capacity to perform the delayed response task 
produced by the receptor blockade (Krystal {\it et al.,} 1994; Adler {\it et al.,} 1998, 
Pontecorvo {\it et al.,} 1991; Cole {\it et al.,} 1993; Aura {\it et al.,} 1999).
Similarly the injection of the NMDA receptor antagonists brings about weakening of the delayed activity 
demonstrated in the electrophysiological studies (Javitt {et al.,} 1996, Dudkin {\it et al.,} 1997a). 
At the same time intracortical perfusion with glutamate receptor agonists  
both improves the performance in the delayed response task and 
increases the duration of WM information storage (Dudkin {\it et al.,} 1997a and b).
Such evidence is especially interesting since 
administration of NMDA receptor antagonists (PCP or ketamine)
reproduces many of the symptoms of schizophrenia including the 
deficits in WM.

Out of many unusual properties of NMDA channel 
two may seem to be essential for the memory storage
purposes: the non-linearity of its current-voltage characteristics
and high affinity to the neurotransmitter, resulting in long lasting
excitatory post synaptic current (EPSC).
The former have been recently implicated in being crucial for WM (Lisman {\it et al.,} 1998). 
It was shown that by carefully balancing NMDA, AMPA, and GABA 
synaptic currents one can produce an N-shaped synaptic current-voltage characteristic, which  
complemented by the recurrent neural circuitry results in the bistability. 
This approach implies therefore that a single neuron can be bistable
and maintain a stable membrane state corresponding to high or low firing frequencies (Camperi and Wang, 1998).
Such a bistability therefore is extremely fragile and can be easily destroyed by 
disturbing the balance between NMDA, GABA or AMPA conductances. 
This may occur in the experiments with the NMDA antagonist.

The single neuron bistability should be contrasted to a more conventional paradigm, in which the
bistability originates solely from the network feedback.
It is based on the non-linear relation between the synaptic currents
entering the neuron and its average firing rate 
(Amit and Tsodyks, 1991a, b; Amit and Brunel, 1997). 
Such a mechanism is not sensitive to the membrane voltage and therefore can be realized using e.g. AMPA receptor.
Thus it may seem that this mechanism is ruled out by the NMDA antagonist experiments.     

In this paper we study how the properties of the synaptic receptor can influence the 
behavior of the conventional bistable network. We therefore 
disregard the phenomena associated with the non-linearity of synaptic
current-voltage relation (Lisman {\it et al.,} 1998). In effect we study the deviations from
the mean-field picture suggested earlier (see e. g. Amit and Tsodyks, 1991a, b).
The main result of this paper concerns the question of stability
of the high frequency attractor. This state is locally stable in the mean-field picture,
i. e. small synaptic and other noises cannot kick the system out of the 
basin of attraction surrounding the state. However this state is not guaranteed to be globally 
stable. After a certain period of time a large fluctuation of the 
synaptic currents makes the system reach the edge of the attraction basin and
the persistent memory state decay into the low frequency state.
We therefore find how long can the working memory be maintained, subject to 
the influence of noises.

Before the network reaches the edge of attraction basin it 
performs an excursion into the region of parameters which
is rarely visited in usual circumstances. 
This justifies the use of term ``instanton'' to name such an excursion,
emphasizing the analogy with the particle traveling in the classically
forbidden region in quantum mechanics.
Such analogy have been used before in application 
to the perception of ambiguous signals by Bialek and DeWeese, 1995.
The present study however is based on microscopic picture,
deriving the decay times from the synaptic properties.   

In the most trivial scenario the network can shut itself 
down by not releasing neurotransmitter in a certain fraction 
of synapses of {\em all} neurons. Thus all the 
neurons cross the border of attraction basin simultaneously.
This mechanism of memory state decay is shown to be ineffective.
Instead the network chooses to cross the border by {\em small groups} 
of neurons, taking advantage of large combinatorial
space spanned by such groups. The cooperative nature of decay in this problem 
is analogous to some examples of tunneling of macroscopic objects in condensed matter physics 
(Lee and Larkin, 1978; Levitov {\it et al.,} 1995). 

The decay time of the memory state evaluated below {\it increases} 
with the size of the network growing (see Section~\ref{GLOBAL_STABILITY}). 
This is a natural result
since in a large network the relative strength of noise 
is small according to the central limit theorem.
Therefore if one is given the minimum time during which the
information has to be stored, there is the minimum
size of the network that can perform this task.
We calculate therefore the minimum number of neurons
necessary to store one bit of information in a recurrent network.
This number weakly depends on the storage time and
for the majority of realistic cases is equal to 5-15.

Two complimentary measures of WM stability, 
the memory decay time and the minimum number of neurons,
necessary to store one bit, are shown below to 
depend on the synaptic channel properties.
The former grows exponentially with increasing
duration of EPSC, which is unusually large for the 
NMDA channel due to high affinity to glutamate.
This allows to interpret the NMDA channel blocking
experiments in the framework of conventional network bistability.

In recent study Wang, 1999 considered a similar problem 
for the network subjected to the influence of {\em external} noises,
disregarding the effects of finiteness of the probability of neurotrasmitter release. 
In our work we analyze the effects of synaptic failures
on the stability of the persistent state. We therefore
consider the {\em internal} sources of noise.
Our study is therefore complimentary to Wang, 1999.

Analytical calculations presented below are confirmed by computer simulations.
For the individual neurons we use the modified leaky integrate-and-fire model due to Stevens and
Zador, 1998, which is shown to reproduce correctly the timing of spikes {\it in vitro}.
Such realistic neurons may have firing frequencies within the range 
15-30Hz for purely excitatory network in the absence of inhibitory inputs (Section~\ref{MEAN_FIELD}). 
This solves the high firing frequency problem (see e.g. Amit and Tsodyks, 1991a). 
The resolution of the problem is based on the unusual property of the
Stevens and Zador neuron, which in contrast to the simple leaky
integrator has {\em two} time scales. These are the membrane time constant
and characteristic time during which the time-constant changes.
The latter, being much longer than the former, determines
the minimum firing frequency in the recurrent network, making
it consistent to the physiologically observed values (see Section~\ref{MEAN_FIELD}
for more detail).


\section{DEFINITION OF THE MODEL AND ITS MEAN-FIELD SOLUTION}
\label{MEAN_FIELD}

In this section we first examine the properties of single neuron,
define our network model, and, finally solve the network approximately, using the
mean-field approximation.

The Stevens-Zador (SZ) model of neuron is an extension of the standard 
leaky integrator (see e.g. Tuckwell, 1998). It is shown to accurately predict the spike timings in layer 2/3 cells
of rat sensory neocortex. The membrane potential $V$
satisfies the leaky integrator equation with time-varying resting potential $E$ and 
integration time $\tau$:
\begin{equation}
\dot{V} = \frac{E\left(t\right) - V}{\tau\left(t\right)} + I(t).
\label{SZequation}
\end{equation}
Here $t$ is the time elapsed since the last spike generated, and the input current $I(t)$
is measured in volts per second. 
When the membrane voltage reaches the threshold voltage $\theta$ 
the neuron emits a spike and the voltage is reset to $V_{reset}$.

%
%
\begin{figure}
(a) 

\vspace{0.1in} 
\centerline{
\psfig{file=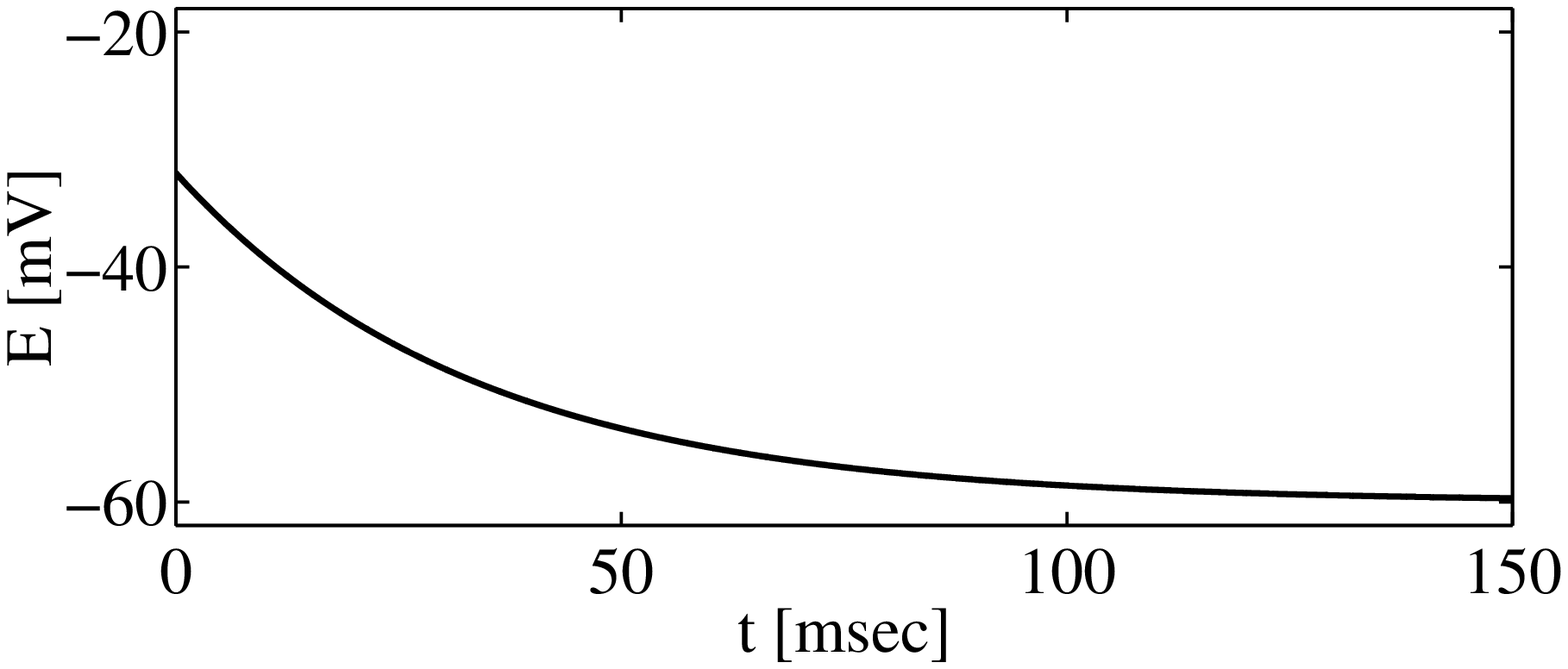,width=3.2in,bbllx=42pt,bblly=195pt,bburx=544pt,bbury=404pt}
}
(b) 

\vspace{0.1in} 
\centerline{
\psfig{file=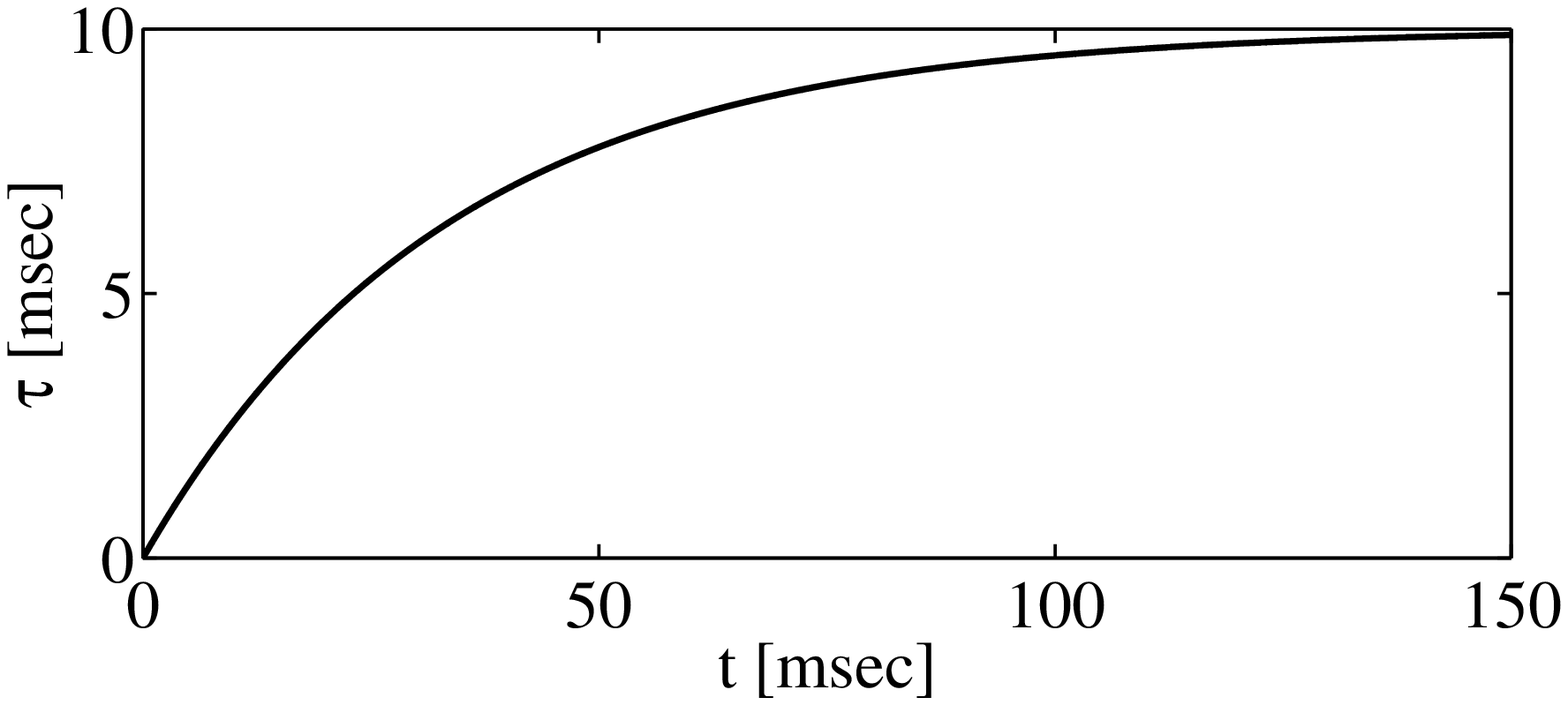,width=3.2in,bbllx=42pt,bblly=195pt,bburx=544pt,bbury=404pt}
}
\vspace{0.1in} 
\caption{
The somatic membrane resting potential (a) and the integration time (b) 
as functions of the time elapsed since the last spike. 
\label{fig10}
}
\end{figure}

This model is quite general to describe many
types of neurons, differing only by the functions $E(t)$, $\tau(t)$, and
the parameters $\theta$ and $V_{reset}$. For the pyramidal cells in cortical layer 2/3
of rats the functions can be fitted by
\begin{equation}
E\left(t\right) = E_0 - \Delta E \left[1 - \exp\left( - \alpha t/\tau_0 \right)\right]
\label{SZE}
\end{equation}
and
\begin{equation}
\tau\left(t\right) = \tau_0 \left[1 - \exp\left( - \alpha t/\tau_0 \right)\right].
\label{SZtau}
\end{equation}
The parameters of the model for these cells have the following numerical values:
$E_0 = -32$mV, $\Delta E = 28$mV, $\alpha = 0.3$, $\tau_0=10$msec, $\theta=-22$mV, and $V_{reset}=-32$mV
(Stevens and Zador, 1998). The resting potential and integration time with these 
parameters are shown in Figure~\ref{fig10}.
\footnote{Since the time constant is zero at $t=0$ 
to resolve the singularity an implicit Runge-Kutta scheme should be used 
in numerical integration of \protect{(\ref{SZequation})}. }

In the next step we calculate the transduction function $f=T(I)$, relating the average 
external current to the average firing frequency. This function 
is evaluated in Appendix~\ref{AppendixA} and is shown in Figure~\ref{fig20}.   
The closed hand expression for this function cannot be obtained. 
Some approximate asymptotic expressions can be found however. 
For large frequencies ($f \gg \alpha/\tau_0 = 30Hz$) it is
approximately given by the linear function (dashed line in Figure~\ref{fig20}):
\begin{equation}
T(I) \approx g \left(I-I_0\right),
\label{large_frequencies}
\end{equation}
where
\begin{equation}
g = \frac{\alpha}{1+\alpha}\frac{1}{\left|E_0 - \theta\right|},
\label{g}
\end{equation}
and
\begin{equation}
I_0 = \frac{\Delta E}{\tau_0}+\frac{\left|E_0 - \theta\right|}{2\tau_0}\frac{1+\alpha}{1+2\alpha}
\label{I_0}
\end{equation}
For small frequencies ($f \ll \alpha/\tau_0$) we obtain:
\begin{equation}
T(I) \approx\frac{\alpha}{\tau_0} \frac{1}{\ln\left[ \left(\left|E_0 - \theta\right|\right)/
\left( \left|E_0 - \theta\right| + \Delta E - \tau_0 I\right) \right]}
\label{small_frequencies}
\end{equation}
The neuron therefore starts firing significantly when current exceeds the critical value
\begin{equation}
I^* = \frac{\left|E_0 - \theta\right| + \Delta E}{\tau_0}.
\label{I_crit}
\end{equation}
We ignore spontaneous activity in our consideration assuming all firing frequencies below $5 Hz$ to be zero. 
We also disregard the effects related to refractory period since they are irrelevant at frequencies $10-50Hz$.
%
%
\begin{figure}
\centerline{
\psfig{file=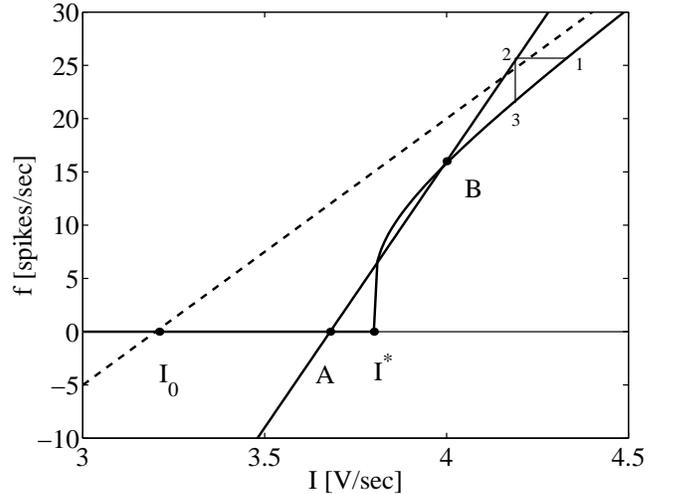,width=3.2in,bbllx=42pt,bblly=195pt,bburx=544pt,bbury=585pt}
}
\vspace{0.1in} 
\caption{
Neuronal transduction function relating the firing frequency and the input current
(solid line \protect{$I_0AI^*B$}). The straight line \protect{$AB$}
describes the network feedback. The dashed line represents the 
asymptotes of the transduction function for large frequencies.  
It is given by \protect{Eq.~(\ref{large_frequencies})}. The expressions for \protect{$I_0$ and $I^*$}
are provided by \protect{Eqs.~(\ref{I_0}) and (\ref{I_crit})}. The points $A$ and $B$ are the low and high 
frequency attractors respectively. The numbers $123$ demonstrate the trajectory
of the system approaching the high frequency attractor.
\label{fig20}
}
\end{figure}

We consider the network consisting of $N$ SZ neurons, establishing all-to-all connections.
After each neuron emits a spike a EPSC is generated in the input currents of all cells
with probability $\kappa$. This is intended to simulate the finiteness of probability of the
neurotransmitter release in synapse. The total input current of $n$-th neuron is therefore
\begin{equation}
I_n(t) = \sum_{k=1}^N \sum{s_k} b_{n,s_k} j(t-t_{s_k}) + I_n^{ext}(t),
\label{totalEPSC}
\end{equation}
where $k$ enumerates the neurons making synapses on the $n$-th neuron
(all neurons in the network), $t_{s_k}$ is the time of spike number $s_k$ emitted by
cell $k$, and $I_n^{ext}$ is the external current.
$b$ is a boolean variable equal to $1$ with probability $\kappa$ 
(in our computer simulations always equal to 0.3).
It is the presence of this variable that distinguishes our approach
from Wang, 1999.
We chose the EPSC represented by $j(t)$ to be (see Amit and Brunel, 1997)
\begin{equation}
j = j_0\exp(-t/\tau_{_{\rm EPSC}})H(t),
\label{EPSC}
\end{equation}
where $H(t) = 1$, if $t>0$, and $H(t) = 0$ otherwise. As evident from (\ref{EPSC}) $\tau_{_{\rm EPSC}}$ is the duration of 
EPSC. It is therefore the central variable in our consideration. 


If the number of neurons in the network is large its dynimics is well described
by the mean-field approximation. How large the number of neurons should be is discussed in the next Section. 
For the purposes of mean-field treatment it is sufficient 
to use the average firing frequency $\bar{f}$ and the average input current $\bar{I}$
to describe the network completely. Thus the network dynamics can be 
approximated by one ``effective'' neuron receiving the average input current and firing at the average frequency. 
Assume that this hypothetical neuron emits spikes at frequency $f(t)$ (point 1 in Figure~\ref{fig20}). 
Due to the network feedback this results in the input current equal 
to the average of Eq.~(\ref{totalEPSC}), displaced in time by the 
average duration of EPSC, i.e. 
\begin{equation}
I(t+\tau_{_{\rm EPSC}}) \approx \bar{I} = N \kappa j_0 \tau_{_{\rm EPSC}} \bar{f}(t) + \bar{I}^{ext}.
\label{MFcurrent}
\end{equation}
This corresponds to the transition between points 1 and 2 in Figure~\ref{fig20}.
At last the transition between the input current and the firing frequency is accomplished
by the transduction function $T(I)$ (points 2 and 3). The delay due to this
transition is of the order of somatic membrane time constant ($\sim 10$msec) and is
negligible compaed to $\tau_{_{\rm EPSC}}\sim 100$msec. We therefore obtain the equation
on the firing frequency $\bar{f}(t+\tau_{_{\rm EPSC}}) = T\left( \bar{I}(\bar{f}(t)) \right)$
(Wilson and Cowan, 1972), or using the Taylor expansion 
\begin{equation}
\tau_{_{\rm EPSC}}\dot{\bar{f}} = T(\bar{I}(\bar{f})) - \bar{f}.
\label{MFequation}
\end{equation}

To obtain the steady state solutions we set the time derivatives in (\ref{MFequation}) to zero
\begin{equation}
T(I) = I^{-1}(I).
\end{equation}
Here $I^{-1}(I)$ is the function inverse to (\ref{MFcurrent}). It is shown in Figure~\ref{fig20}
by the straight solid line. This equation has three solutions, 
two of which are stable. They are marked in by letters $A$ and $B$ in the Figure. 

Point $B$ represents the high frequency attractor. 
Frequencies obtained in SZ model neurons are not too high. 
They are of the order of $\alpha/\tau_0$, i.e. in the range $15-30$ Hz.
This coinsides with the range of frequencies observed in the delayed activity
experiments (Miyashita, 1988; Miyashita and Chang, 1988; Sakai and Miyashita, 1991; Funahashi {\it et al.,} 1989;
Goldman-Rakic {\it et al.,} 1990; Fuster, 1995). 
The reason for the relatively low firing frequency rate is as follows.
For the leaky integrator model the characteristic firing
rates in the recurrent network are of the order of $1/\tau$,
i. e. are in the range $50-100$Hz. For the SZ neuron $\tau$
is even smaller (see Figure~\ref{fig10}b), and therefore it seems that the firing rates should be 
larger than for leaky intergator. This however is 
no true, since for the SZ neuron, in contrast to the leaky
integrator, there is the second time scale. It is the characteristic
time of variation of the time constant $\tau_0/\alpha \sim 30-50$ msec (see Figure~\ref{fig10}b). 
Since the time constant itself is very short it becomes irrelevant 
for the spike generation purposes at low firing rates and
the second time scale determines the characteristic 
frequency. It is therefore in the range $15-30$Hz.

Finally we would like to discuss the stability of attactor $B$. 
It can be locally stable, i.e. small noises cannot produce the transition from state $B$ to state $A$.
The condition for this follows from the linerized near equilibrium Eq.~(\ref{MFequation}):
\begin{equation}
\nu = \frac{dT}{d\bar{I}} \frac{d\bar{I}}{d\bar{f}} < 1
\label{FeedbackCoefficient}
\end{equation}
Very rarely, however, a large fluctuation of noise can occur, that
kicks the system out of the basin of attraction of state $B$.
It is therefore {\em never} globally stable.
This is the topic of the next Section.


\section{BEYOND MEAN-FIELD APPROXIMATION}
\label{GLOBAL_STABILITY}

\begin{figure}
(a) 

\vspace{0.1in} 
\centerline{
\psfig{file=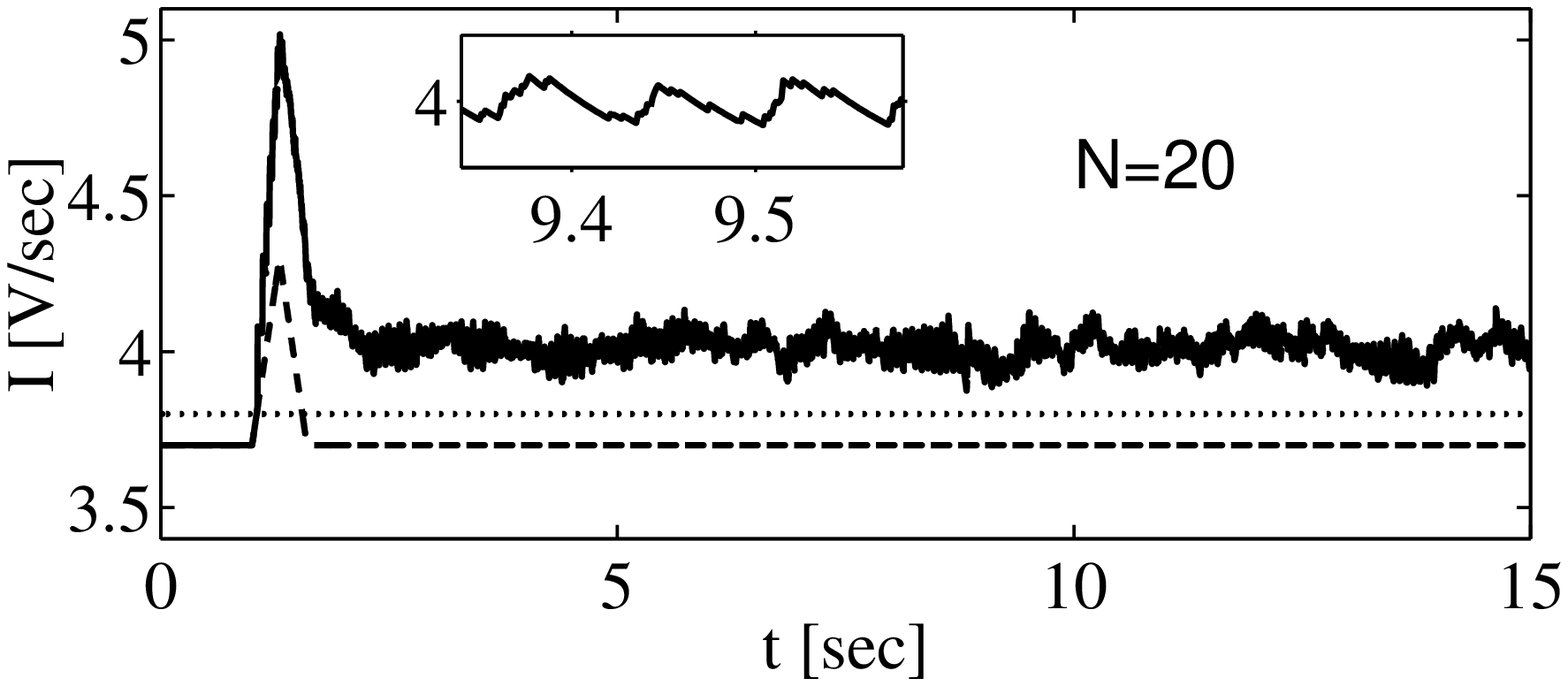,width=3.2in,bbllx=42pt,bblly=195pt,bburx=544pt,bbury=404pt}
}
(b) 

\vspace{0.1in} 
\centerline{
\psfig{file=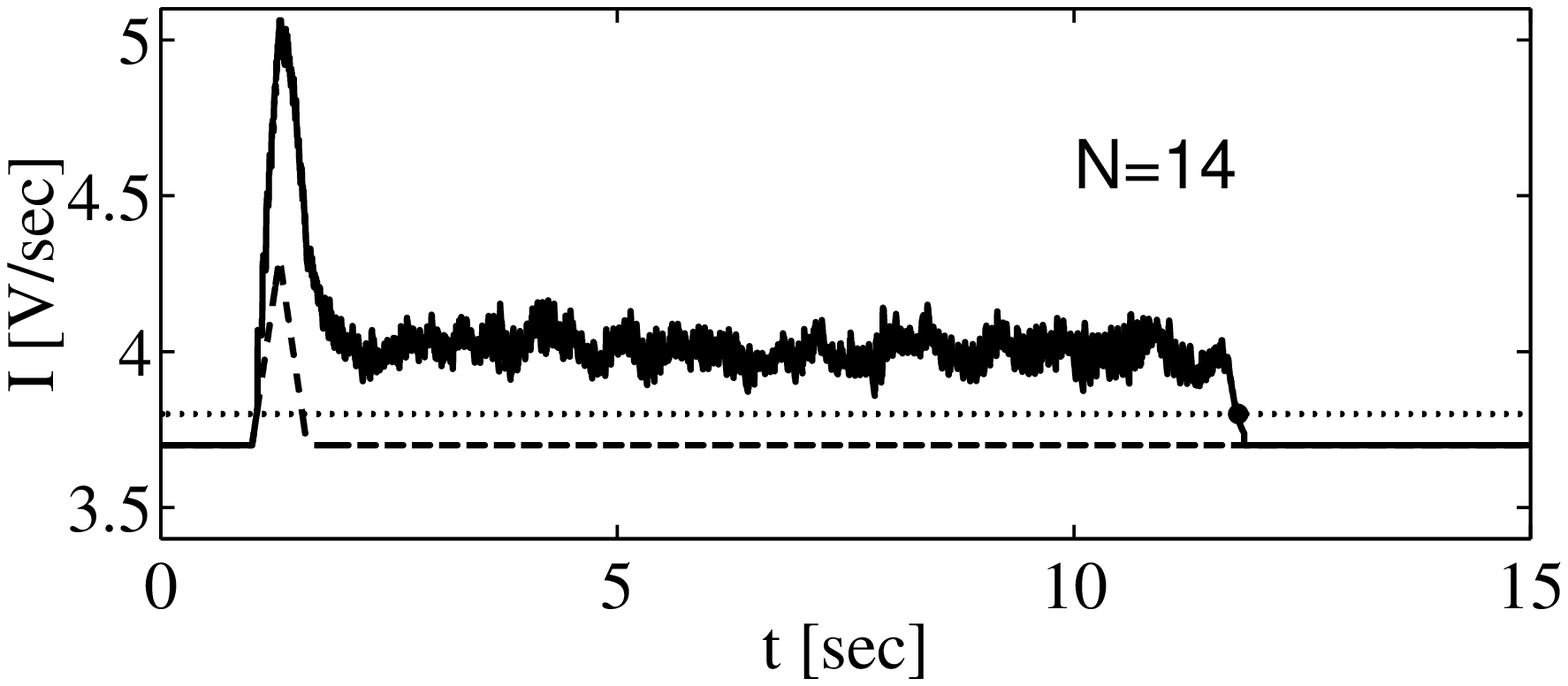,width=3.2in,bbllx=42pt,bblly=195pt,bburx=544pt,bbury=404pt}
}
\vspace{0.1in} 
\caption{ 
Total input current averaged over all the neurons in the network. A short pulse of the 
external current brings about the transition to the high frequency state, in which the
total current is larger than the external current due to the feedback. (a) In large networks (N=20)
the delayed activity can persist virtually forever, until the external inputs are changed.
The inset shows the traces of neuronal synchronization:
the average current experiences oscillations at the average firing frequency.
(b) In small network (N=14) the high frequency state can decay and the delayed activity 
disapear. This results in the average current crossing the edge of the attraction basin,
trace of which is shown by the dotted line. 
\label{fig25}
}
\end{figure}

Computer simulations show that our network can successfully generate
delayed activity response. An example is shown in Figure~\ref{fig25}a
where a short pulse of external current (dashed line) produced transition to
the high frequency state. This state is well described by the mean-field
treatment given in the previous section. 
This is true, however, only for the networks containing a large number of neurons $N$. If the size of the 
network is smaller than some critical number $N^*$ the following phenomenon 
is observed (see Figure~\ref{fig25}b). The fluctuations of the current
reach the edge of the attraction basin (dotted line) and the network
abruptly shuts down, jumping from high frequency to the low frequency state.
The quantitative treatment of these events is the subject of this section.  
%
%

\begin{figure}
\centerline{
\psfig{file=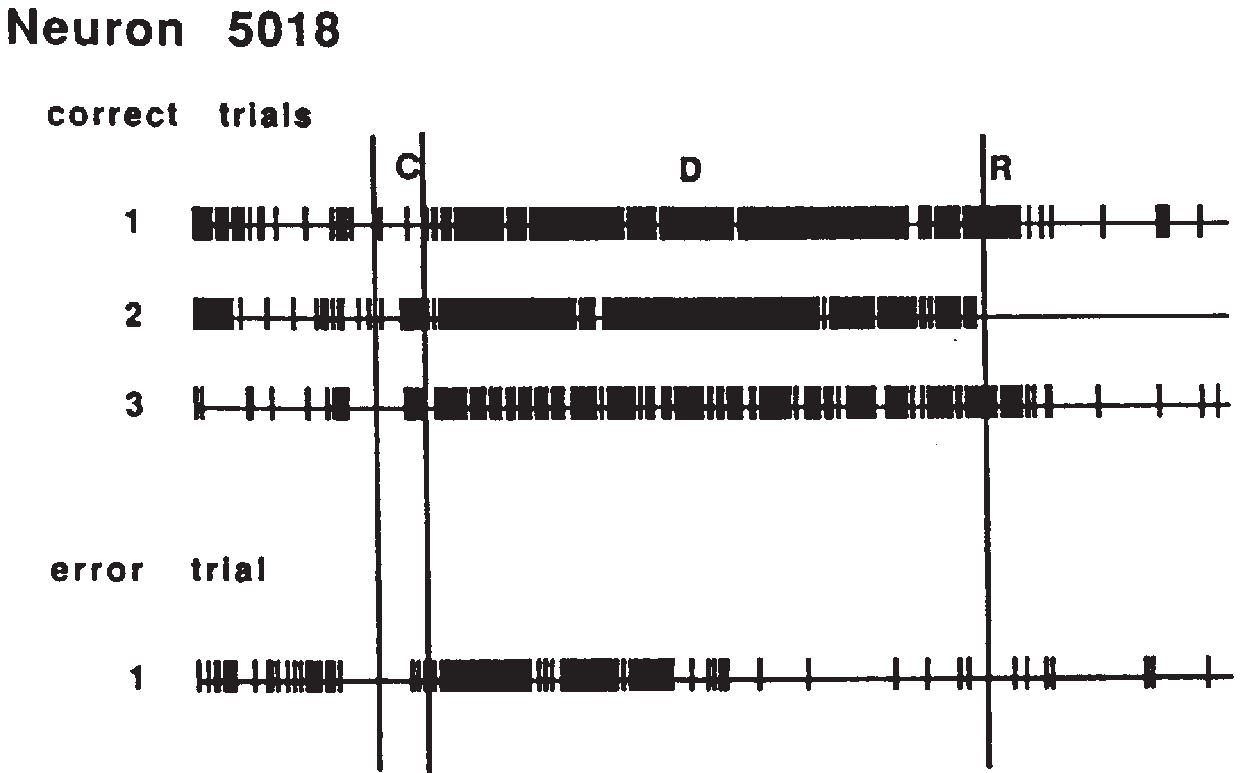,width=3.2in,bbllx=4pt,bblly=0pt,bburx=356pt,bbury=221pt}
}
\vspace{0.1in} 
\caption{Sudden decay of the dalayed activity in the neuron
in prefrontal cortex during error trial (bottom figure) observed
by Funahashi {\it et al.,} 1989.
\label{fig27}
}
\end{figure}

Similar decay processes have been observed by Funahashi {\it et al.,} 1989
in prefrontal cortex. One of the examples of such error trials 
is shown in Figure~\ref{fig27}.
%
%

\begin{figure}
\centerline{
\psfig{file=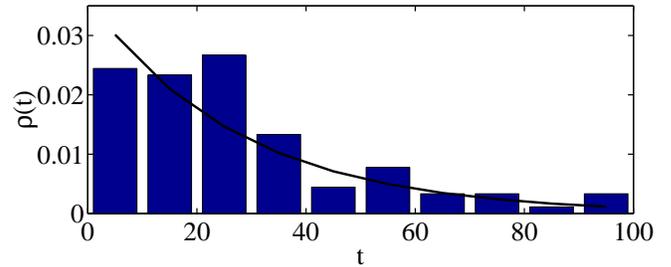,width=3.2in,bbllx=42pt,bblly=195pt,bburx=544pt,bbury=404pt}
}
\vspace{0.1in} 
\caption{ The density of probability of decay at time \protect{$t$}
obtained in the numerical experiment (bars) and the exponential distribution
with \protect{$\bar{t}=28$sec} (solid curve). 
The latter is given by \protect{Eq.~(\ref{Poisson2})}. The numerical results are include 100 trials
with \protect{$\bar{f}\approx 16$Hz, $N=14$, $\tau_{_{\rm EPSC}}=80$msec}.
\label{fig30}
}
\end{figure}

Although the decay is abrupt, the moment at which it occurs is
not reprodusable from experiment to experiment. It is
of interest therefore to study the distribution of the 
time interval between the initiation of the delayed activity and the moment of its decay.
Our simulations and the arguments given in Appendix~\ref{Fluctuations} show that the decay can be considered a Poisson process.
The decay times have therefore an exponential distribution (see Figure~\ref{fig30}).

%
%

\begin{figure}
\centerline{
\psfig{file=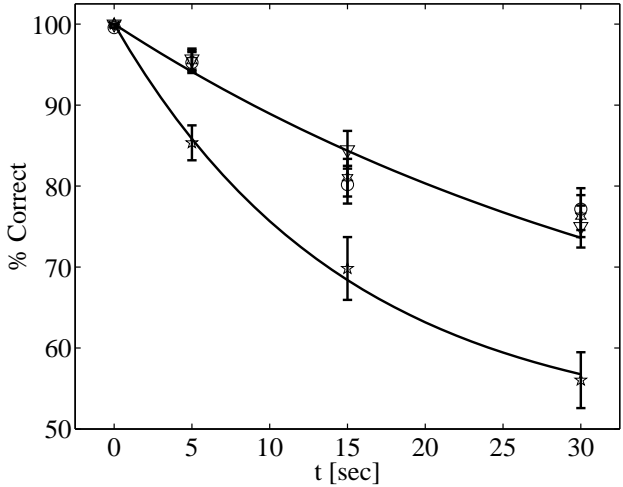,width=3.2in,bbllx=35pt,bblly=184pt,bburx=538pt,bbury=580pt}
}
\vspace{0.1in} 
\caption{ The performance of rats in delayed matching to position task (from Cole {\it et al.,} 1993).
Different markers show various degrees of sedation with competitive NMDA antagonist CPP:
circles 0 mg/kg; triangles 1 mg/kg; hexagons 3 mg/kg; pentagons 10 mg/kg. The solid lines 
represent the fits by \protect{Eq.~(\ref{Correct})}. The upper and lower lines correspond
to the average decay times 40 and 15 seconds respectively.
\label{fig40}
}
\end{figure}

Since failure to maintain the persistent activity entails the loss of the
memory and incorrect performance in the delayed responce task,
at least in monkey's prefrontal cortex (Funahashi {\it et al.}, 1989), one can 
use the conclusion about Poisson distribution to interpret some psychophysical data.
In some experiments on rats performing the binary delayed matching to position tasks
deterioration of WM is observed as a function of delay time (Cole {\it et al.,} 1993).
The detereoration was characterized by ``forgetting'' curve, with matching accuracy 
decreasing from about 100\% at zero delay to approximately 70\% at 30 second delay.
The performance of the animal approaches the regime of random guessing 
with 50\% of correct responses in this binary task (Figure~\ref{fig40}).
Our prediction for the shape of the ``forgetting'' curve that follows
from the Poisson distribution of delayed activity times is
\begin{equation}
Correct = (1+\exp(-t/\bar{t}))/2\times 100\%.
\label{Correct}
\end{equation}
This prediction is used to fit the experimental data in Figure~\ref{fig40}.
%
%
This Figure also shows the effect of competitive NMDA antagonist CPP. 
Application of the antagonist reduces the average memory retention time $\bar{t}$ from
about 40 seconds to 15 seconds. The presence of non-compatitive 
antagonists impairs performance even at zero delay (Cole {\it et al.,} 1993; Pontecorvo{\it et al.,} 1991). 
Non-competitive antagonists have therefore an effect on the components of animal behavior 
different from WM. When the delayed component of ``forgetting'' curve is extracted
it can be well fitted by an expression containing exponential similar to (\ref{Correct}). 
Such fits also show that the average WM storage time $\bar{t}$ decreases with application of NMDA antagonist.

Our calculation in Appendix~\ref{Fluctuations} show that the average memory storage time
is given by
\begin{equation}
\bar{t} \sim \tau_{_{\rm EPSC}} \exp \left[{ \kappa f\tau_{_{\rm EPSC}} N^2 \left( \Delta I / I \right)^3 }\right].
\label{AverageTime}
\end{equation}
Here $\Delta I$ is the distance to the edge of the attraction basin 
from the stable state and $I$ is the average feedback current.
This result holds if $f\tau_{_{\rm EPSC}} \gg 1$.
Because $\bar{t}$ is of the order of tens of seconds and $\tau_{_{\rm EPSC}}$ 
is approximately $100$ msec the exponential in (\ref{AverageTime}) is of the order of $10^2$-$10^3$
for the realistic cases.

There are therefore two ways how synaptic receptor blockade can affect $\bar{t}$.
First, the attenuation of the EPSC decreases the average firing frequency $f$.
Second, it moves the system closer to the edge of the attraction basin, reducing $\Delta I$.
Both factors increase the effect of noise onto the system,
decreasing the average memory storage time.
This is manifested by Eq.~(\ref{AverageTime}).

Another consequence of the formula is the importance of NMDA receptor
for the WM storage. It is based on the large affinity of the receptor
to glutamate, leading to long EPSC ($\tau_{_{\rm NMDA}}\approx 100$ msec),
compared for example to AMPA receptor ($\tau_{_{\rm AMPA}}\approx 15$ msec).
Eq.~(\ref{AverageTime}) implies that if AMPA receptor is used 
in the bistable neural net and all other parameters ($\kappa$, $f$, $N$, and $\Delta I/I$) are kept the 
the same, the memory storage time is equal to $\tau_{\rm AMPA}\approx 15$ msec.
Thus it is not suprising that NMDA receptor is chosen by evolution as a mediator
of WM and the highest density of the receptor is observed in places 
involved into the WM storage, i.e. in prefrontal cortex (Cotman {\it et al.,} 1987).
%
%
\begin{figure}
\centerline{
\psfig{file=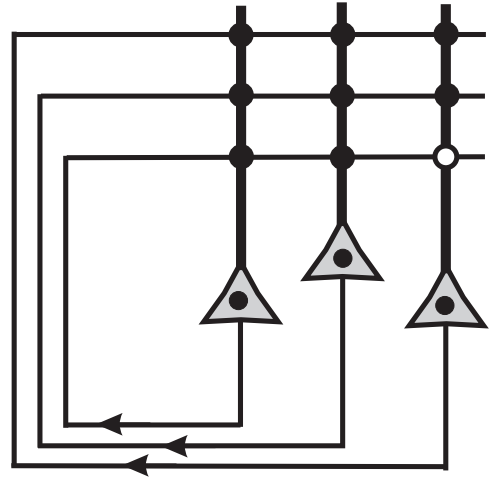,width=2.5in,bbllx=194pt,bblly=285pt,bburx=383pt,bbury=481pt}
}
\vspace{0.1in} 
\caption{ The network used in text to illustrate the optimum fluctuation leading
to the delayed activity decay. 
The synapses are shown by circles. 
The synapses releasing neurotransmitter are
shown by full circles. The failing synapse is shown by the open circle. 
The dendrites and axons are represented by thick and thin lines respectively.
\label{fig50}
}
\end{figure}
We now would like to illustrate what processes lead to the decay time given by (\ref{AverageTime}).
Consider a simple network consisting of three neurons (Figure~\ref{fig50}).
Assume that the ratio $\Delta I/I$ for this network is equal to 1/3.
This implies that the neurons have to lose only 1/3 of their recurrent input 
current due to a fluctuation to stop firing. This can be accomplished by various means.
Our research shows that the most effective fluctuation is as follows.
Due to probabilistic nature of synaptic transmition some of the 
synapses release glutamate when spike arrives onto the presinaptic terminal
(full circles in Figure~\ref{fig50}) some fail to do so (open circle). 
It is easy to see that if the same synapse 
fails to release neurotransmitter in responce to any spike arriving during the time
interval $\tau_{_{\rm EPSC}}$ the reverberations of current terminate.
Indeed, the failing synapse (open circle) deprives the neuron of 1/3 of its current.
This is just enough to put the input current into the neuron below the threshold.
The neuron therefore stops firing. This deprives the entire network, consisting of three neurons, of 
1/3 of its feedback current. Therefore the delayed activity in this network terminates.

In the the most reasonable alternative mechanism of decay the average mean-field current
would reach the edge of the attraction basin i.e. would be reduced by 1/3. This can be
accomplished by shutting down three synapses in three different neurons instead of one. Such mechanism is therefore 
less effective than the proposed above. Quantitatively the difference 
is manifested in reducing the exponent of the factor containg $\Delta I / I < 1$ in Eq.~(\ref{AverageTime})
from 3 to 2. This bring about an {\it increase} in the average storage time.
The mean-field mechanism is therefore less restrictive than the proposed one
and is disregarded in this paper.

\begin{figure}
\centerline{
\psfig{file=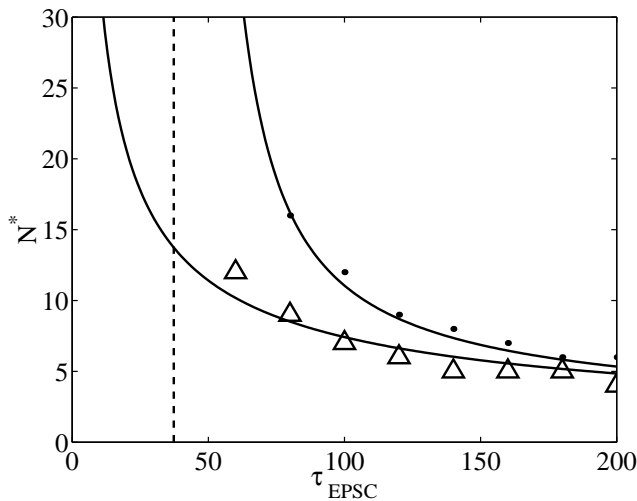,width=3.2in,bbllx=35pt,bblly=184pt,bburx=538pt,bbury=580pt}
}
\vspace{0.1in} 
\caption{ 
The minimum number of neurons \protect{$N^*$} 
needed to maintain delayed activity with average decay time 20 sec
for \protect{$\bar{f}\approx 16$Hz}.
The dots show computational results for the case of no noise in the
external inputs. The triangles correspond to the case
of 10\% white noise added to the external current. 
The solid lines represent the result of analytical calculation
described below in the text. The dashed line 
is the vertical asymptote of \protect{$N^*$}
for the case of no external noise.
No delayed activity can exist 
to the left of this line, i.e.
for the synaptic receptor time constant smaller than
37 msec. 
\label{fig52}
}
\end{figure}

Since an increase of the number of neurons in the network $N$ 
dramatically influences the memory storage time according to Eq.~(\ref{AverageTime}),
another characteristic of the reliability of WM circuit is the minimum number of 
neurons $N^*$ necessary to store one bit of information during time $\bar{t}$. 
We first study this quantity computationally as follows. We run the 
network simulation many times with the same values of $N$ and $\tau_{_{\rm EPSC}}$.
We then determine the average decay time $\bar{t}$. Having done this
we decrease or increase $N$ depending on wheater $\bar{t}$ is larger or smaller
than a given value (20 sec in all our simulations). This process
converges to the number of neurons necessary to sustain the delayed activity $N^*$
for the given value of $\tau_{_{\rm EPSC}}$. The process is then repeated 
for different values of synaptic time-constant. However, the
network feedback is always renormalized so that the firing frequency 
stays the same, close to the physiologically feasible
value $\bar{f}\approx 16Hz$. This corresponds to the attractor state
shown in Figure~(\ref{fig20}). The resulting dependence of $N^*$ 
versus $\tau_{_{\rm EPSC}}$ is shown in Figure~\ref{fig52} by markers.

%
%

There are two sets of computational results in Figure~\ref{fig52}. 
The higher values of $N^*$ are obtained for the network with
no noise in the external inputs (dots in Figure~\ref{fig52}). 
The only source of noise in such a
network is therefore the unreliability of synaptic connections.
It appears for this case that {\em no} delayed activity can exist 
for $\tau_{_{\rm EPSC}}$ below 37 msec.
The lower values of $N^*$ are obtained for the case 
of white noise added to the external input (triangles). 
The amplitute of white noise is $10\%$ of the total input current
and the correlation time is 1 msec. The limiting value of synaptic 
time constant for which the dalayed activity is not possible is
much smaller for this case ($\approx 5$ msec).

This result may seem counter-intuitive. Having added the external
noise we increased the viability of the high frequency state,
reducing the effects of the internal synaptic noise.
This however is not so suprising if one takes the
neuronal synchrony into account. Synchrony
of neuronal firing results in oscillations 
in the average input current (see inset in Figure~\ref{fig20}a).
Such oscillations periodically bring the system closer to the 
edge of the attraction basin, creating additional 
opprotunities for decay. Thus the syncronous network is
less stable than the asynchronous one. 
External noice attenuates neuronal synchrony, smearing the 
ascillations of the average input current. 
Thus the system with noise in the external inputs should have
a larger decay time and smaller critical number of neurons $N^*$.
This idea is discussed quantitatively below in this Section.
This is similar to the stabilisation of the mean-field solutions in
the networks  of pulse-coupled oscillators (Abbott and Vreeswijk, 1993).

%
%
\begin{figure}
\centerline{
\psfig{file=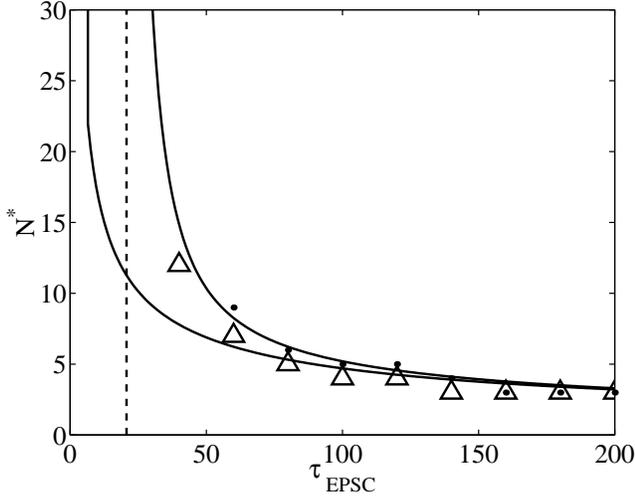,width=3.2in,bbllx=35pt,bblly=184pt,bburx=538pt,bbury=580pt}
}
\vspace{0.1in} 
\caption{ 
The minimum number of neurons \protect{$N^*$} 
needed to maintain delayed activity with average decay time 20 sec 
for the attaractor with average firing frequency 28 Hz.
Notations are the same as in Figure~\protect{\ref{fig52}}.
The cut-off value of synaptic time constant is 27 msec.
The numerical results with external noise in this case
are slightly above the theoretical prediction due to
insufficient suppression of synchrony.
\label{fig54}
}
\end{figure}

Figure~\ref{fig54} shows the results of similar calculations 
for the attractor state with a higher average firing frequency ($\bar{f}=28$Hz).
This network shows higher reliability and smaller values of
$N^*$ for both synchronous and asyncronous (10\% noise added) regimes.
This is consistent with Eq.~(\ref{AverageTime}). The results
of analytical calculations (solid lines) show satisfactory 
agreement with the computer modeling (markers). The rest of the 
Section is dedicated to the discussion of different aspects of the analytical
calculations and their results. A more thorough treatment 
of the problem can be found in Appendix~\ref{Fluctuations}.

To evaluate the minimum number of neurons in the closed hand form we 
solve Eq.~(\ref{AverageTime}) for $N$
\begin{equation}
N^* \approx N_1^* \equiv \frac {1} {\sqrt{\kappa f \tau_{_{\rm EPSC}}}} \left( \frac{I}{\Delta I}\right)^{3/2} \sqrt{ \log\left( \frac{\bar{t}} {\tau_{_{\rm EPSC}}}\right) }
\label{Nmin1}
\end{equation}
This equation is valid for $f\tau_{_{\rm EPSC}} \gg 1$. In the opposite case
$f\tau_{_{\rm EPSC}} \lesssim 1$ it has to be amended. To obtain the 
correct expression in the latter case the following considerations should be taken into account. 

\subsection{Fluctuations of the average mean-field current.}

It is obvious that when the mean-field attractor becomes less and less stable in
the local sense, i.e. the dimensionless feedback coefficient $\nu < 1$ [Eq.(\ref{FeedbackCoefficient})] 
approaches unity, the global stability should also suffer.
Indeed, if the system is weakly locally stable the fluctuations in the average current
are large. This should facilitate global instability.
The facilitation can be accounted for by noticing that the transition
from high frequency state $B$ to the low frequency one $A$ is 
most probable when the average current is low. Hence to obtain the most realistic probability of
transition and correct values for $\bar{t}$ and $N^*$ one has to decrease 
the values of average current and distance to the edge of the attraction basin by
\begin{equation}
\begin{array}{l}
{\displaystyle I \rightarrow I - \delta I,} \\ \\
{\displaystyle \Delta I \rightarrow \Delta I - \delta I,}
\end{array}
\label{Renormalization}
\end{equation}
where the standard deviation of the average current $\delta I$ is calculated in Appendix~\ref{Fluctuations}
\begin{equation}
\delta I = \frac {I} {N\sqrt{2(1-\nu)\kappa f \tau_{_{\rm EPSC}}}}.
\label{deltaI}
\end{equation}
Here $I$ is the average current before the shift.
This correction decreases the ratio $\Delta I/I$, decreasing $\bar{t}$ [see Eq.~(\ref{AverageTime})]. 
This implies the reduced reliability of the network due to the average current fluctuations.

\subsection{Large combinatorial space covered by small groups of neurons attempting to cross
the attraction basin edge.}

In the simple example with three neurons, in principle, each of them can contribute to the 
process of decay. If the network is larger the number of potentially dangerous 
groups grows as a binomial coefficient $C_N^n$, where $n = N\Delta I/I \leq N$ is the size of  
the dangerous group. The correction to $N^*$ can be expressed as follows (Appendix~\ref{Fluctuations}):
\begin{equation}
N^* = \sqrt{\left( N_1^* \right)^2 + \left( N_2^* \right)^2} + N_2^*.
\label{Nmin}
\end{equation}
Here $N_2^*$ is the combinatorial correction
\begin{equation}
N_2^* = \left( \frac{I}{\Delta I} \right)^3 \frac{\phi\left( \Delta I/I \right) + \left( \Delta I/I \right) \ln A}
{\kappa f \tau_{_{\rm EPSC}}}.
\label{Nmin2}
\end{equation}
Here
\begin{equation}
\begin{array}{l}
{\displaystyle A = \frac{I}{\Delta I \sqrt{4\pi N^* \kappa f \tau_{_{\rm EPSC}}}},} \\ \\
{\displaystyle \phi(x) = x\ln \left[1/x\right] + (1-x) \ln \left[1/(1-x)\right].}
\end{array}
\end{equation}
Since the combinatorial contribution $N_2^*$ is proportional to $1/f \tau_{_{\rm EPSC}}$ it is 
negligible compared to $N_1^* \propto 1/\sqrt{f \tau_{_{\rm EPSC}}}$ at large values of $\tau_{_{\rm EPSC}}$.
Therefore $N^* \approx N_1^*$ as claimed by Eq.~(\ref{Nmin1}) in this limit. 
On the other hand if $\tau_{_{\rm EPSC}}$ is small the main contribution to
the critical number of neurons $N^*$ comes from $N^*_2$. 

The right hand sides of Eqs.~(\ref{Nmin1}) and {\ref{Nmin2}} contains
dependence on $N^*$ through the shift in distance to the attraction edge
$\delta I$ and the coefficient $A$. This dependence is weak however
since the former represents a very small correction and the latter
depends on $N^*$ only logarithmically. Nevertheless to generate a numerically
precise prediction we iterate Eq.~(\ref{Nmin1}), (\ref{Nmin}), and (\ref{Nmin1})
until a consistent value of $N^*$ is reached. The results
are shown by lower solid lines in Figures~\ref{fig52} and \ref{fig54}.
They are in a good agreement with numerical simulations in which
the synchrony of firing was suppressed by external noise.

\subsection{Neuronal synchrony.}

The synchronizaton of neuronal firing can be critical for the stability of the 
high frequency attractor. Consider the following simple example (Wang, 1999).
Consider the network consisting on only {\it one} neuron. 
Assume that the external input current to the neuron exceeds the firing threshold 
$I^*$ at certain moment and the neuron starts emiting spikes. 
Assume that the external current is then decreased to some value below the threshold. 
The only possibility for the neuron to keep firing is to receive enough feedback current from itself, so that the
{\em total} input current is above threshfold. 
Suppose then for illustration purposes that the synapse that the neuron forms on itself (autapse) 
is infinitely reliable, i.e. every spike arriving on the presynaptic terminal brings about the release of neurotransmitter.
It is easy to see that if the duration of EPSC is much shorter than the interspike interval 
the persistent activity is impossible. 
Indeed, if EPSC is short, at the moment, when the new spike has to be generated
there is almost no residual EPSC. The input current at that moment is almost entirely due to external
sources, i.e. is below the threshold. 
The new spike generation is therefore impossible.
We conclude that the persistent activity is impossible if EPSC is short, i. e. for 
$f\tau_{_{\rm EPSC}} \ll 1$. Note, that this conclusion is made assuming {\em no noise} in the
system. 

The natural question is why this conclusion is relevant to the network consisting of many
neurons? Our computer modeling shows that in the large network the neuronal firing
pattern is highly sychronized on the scales of the order of $T=1/f$. In essence  
all neurons fire simultaneously and periodically with the period $T$. 
Therefore they behave as one neuron. Hence the delayed activity in such
network is impossible if $f\tau_{_{\rm EPSC}} \ll 1$ even in the absence of
synatic noise. Thus the synchrony facilitates the decay of the delayed activity
brought about by the noise. This is consistent with the results of computer
simulations presented if Figures~\ref{fig52} and \ref{fig54}.

As it was mentioned above synchrony facilitates the instability
by producing the oscillations in the average current 
and therefore by bringing the edge of the attraction basin closer. 
It therefore effectively decreases $\Delta I$.
The magnitude of this decrease is estimated in Appendix~\ref{Fluctuations}
to be 
\begin{equation}
\begin{array}{l}
{\displaystyle \delta I_{syn} \approx \frac {2I}{\pi f\tau_{_{\rm EPSC}}} } \\ \\
{\displaystyle \times  
\frac{ f\tau_{_{\rm EPSC}} - e^{-1/f\tau_{_{\rm EPSC}}}\left( f\tau_{_{\rm EPSC}} + 1 \right)}
{1-e^{-1/f\tau_{_{\rm EPSC}}}}}
\end{array}
\label{deltaIs}
\end{equation}
and
\begin{equation}
\Delta I \rightarrow \Delta I - \delta I_{syn}.
\label{ShiftS}
\end{equation}
Here $I$ is the average current given by (\ref{Renormalization}).
This correction, which is the amplitude of the oscillations of the 
average current due to synchrony, is in good agreement with 
computer modeling (see insert in in Figure~\ref{fig25}a).
The synchrony does not change the average current in the network
significantly. Therefore the latter should not be shifted as in (\ref{Renormalization}).

These corrections imply that if the original value of $\Delta I$ is equal
to the sum $\delta I+\delta I_{syn}$ [see Eq.~(\ref{deltaI})], 
the effective distance to the edge of attraction basin vanishes. 
The delayed activity cannot be sustained under such a condition.
This occurs at small values of $\tau_{_{\rm EPSC}}$ and determines
the positions of vertical asymptotes (dashed lines) in Figures~\ref{fig52} and \ref{fig54}.
This gives a quantitative meaning to the argument of impossibility
of stable delayed activity in synchronous network at small values 
of synaptic time constant given above (see also Wang, 1999).
When the synchrony is diminished by external noise,
the cut off value of $\tau_{_{\rm EPSC}}$ is determined only by $\Delta I = \delta I$
and is therefore much smaller. It is about 5 msec in our computer similations.

Synchrony also affects the values of the critical number 
of neurons for large and small $f\tau_{_{\rm EPSC}}$ ($N^*_1$ and $N^*_2$ respectively)
\begin{equation}
N^*_{1syn} = \left( \frac{I}{\Delta I} \right)^{3/2} \frac{1}{f\tau_{_{\rm EPSC}}}
\sqrt{\frac{2\ln \left( t/\tau_{_{\rm EPSC}}\right) }{\kappa\left( e^{2/f\tau_{_{\rm EPSC}}} -1\right)}},
\label{Nmin1s}
\end{equation}
\begin{equation}
N^*_{2syn} = \left( \frac{I}{\Delta I} \right)^{3} 
\frac{\phi\left( \Delta I/I \right) + \left( \Delta I/I \right) \ln A}
{e^{2/f\tau_{_{\rm EPSC}}} -1} \frac{1}{\kappa f^2 \tau_{_{\rm EPSC}}^2}
\label{Nmin2s}
\end{equation}
These equations are derived in Appendix~\ref{Fluctuations}.
To obtain the value of critical number of neurons $N^*$
Eq.~(\ref{Nmin}) should be used.
Let us compare the latter equations to (\ref{Nmin1}) and (\ref{Nmin2}).
In the limit $f \gg 1$ the expressions for
$N^*_1$ and $N^*_{1syn}$ converge to the same asymptote $N^*_1 \approx N^*_{1syn}$,
whereas both $N^*_2$ and $N^*_{2syn}$ go to zero.

Since again, as in asynchronous case, both $N^*_{1syn}$ and $N^*_{2syn}$
depend on the quantity that we are looking for $N^*$ (however, again, very weakly), 
an iterative procedure should be used to determine the consistent 
value of $N^*$. Having applied this iterative procedure we obtain
the ipper solid lines in Figures~\ref{fig52} and \ref{fig54}.
We therefore obtain an excellent agreement with the results
of the numerical study. In addition we calculate the 
cut-off $\tau_{_{\rm EPSC}}$, below which the delayed activity
is not possible for the synchronous case. For the attractor with
$\bar{f}\approx 16$ Hz the cut-off value is $37$ msec, while 
for the higher frequency attractor ($\bar{f} \approx 28$ Hz)
the value is $21$ msec. These values are shown in Figures~\ref{fig52} and \ref{fig54}
by dashed lines. We therefore conclude that the delayed activity 
mediated by AMPA receptor is impossible in the synchronous case.






\section{DISCUSSION}
\label{DISCUSSION}


In this work we derive the relationship between
the dynamic properties of the synaptic receptor channels
and the stability of the delayed activity. We conclude
that the decay of the latter is a Poisson process with
the average decay time exponentially depending on the 
time constant of EPSC. Our quantitative conclusion applied
to AMPA receptor, having a short EPSC, implies that it is incapable of 
sustaining the persistent activity in case of synchronization
of firing in the network. For the case of asynchronous 
network one needs a large number of neurons $>30$
to store one bit of information with AMPA receptor. On the other
hand NMDA receptor seems to stay away from these 
problems, providing reliable quantum of information
storage with about 15 neurons for both synchronized and asynchronous case.
We therefore suggest an explanation to the obvious from experiments
high significance of the NMDA channel for WM.


One can conclude from our study that if the time constant of NMDA channel EPSC 
is further increased, the WM can be stored for much longer time.
Assume that $\tau_{_{\rm EPSC}}$ is increased by a factor of 2,
for instance, by genetic enhancement (Tang {\it et al.,} 1999).
If the network connectivity, firing frequency, and average currents
stay the same as in the wild type, the 
exponential in Eq.~(\ref{AverageTime}) is increased by a factor of $10^3$.
This implies that the working memory can be stored by such an animal 
for days, instead of minutes. 
Alternatively the the number of neurons responsible for the storage of quantum of information 
can be decreased by a factor $0.7$ keeping the storage time the same.
This implies the higher storage capacity of the brain of mutant animals.


We predict that with normal NMDA receptor the number of neurons 
able to store one bit of information for $40$ sec is about 15. 
Should our theory be applicable to rats performing delayed matching to position task
(Cole {\it et al.,} 1993) the conclusion would be that the 
recurrent circuit responsible for this task contains $<$ 15 neurons.
Of course this would imply that other sources of loss of memory,
such as distraction, are not present.


The use of very simple model network allowed
us to look into the nature of the global instability
of delayed activity. The principal result of this 
paper is that the unreliability of synaptic conductance
provides the most effective channel for the delayed activity
decay. We propose the optimum fluctuation of the
synaptic noises leading to the loss of WM.
Decay rates due to such a fluctuation agree well with
the results of the numerical study (Section~\ref{GLOBAL_STABILITY}).
Thus any theory or computer simulation that does not take the unreliability
of synapses into account does not reproduce
the phenomenon exactly.


We also studied the decay process in the presence of noise in the afferent inputs.
The study suggests that the effect of noise on the WM storage reliability is
not monotonic. Addition of small white noise ($<10$\% of the total external current)
increases the reliability, by destroying synchronization. 
It is therefore beneficial for the WM storage. Further
increase of noise ($>12$\%) destroys WM, producing transitions between 
the low and high frequency states. We conclude therefore
that there is an optimum amount of the afferent noise, which
on one hand smoothes the synchrony out and on the other hand 
does not produce the decay of delayed activity itself.
Further work is needed to study the nature of influence 
of the external noise in the case of unreliable synapses.


If the afferent noise is not too large the neurons fire in synchrony. 
The natural consequences of synchrony are the
decrease of the coefficient of variation of the interspike interval
for single neuron and the increase of the crosscorrelations between neurons.
The latter prediction is consistent with findings of some multielectrode 
studies in monkey prefrontal cortex (see Dudkin {\it et al.,} 1997a).
On the other hand sinchrony may also be relevant to the phenomenon of temporal binding
in the striate cortex (Engel {\it et al.,} 1999; Roskies, 1999).
We argue therefore that temporal binding with the precision
of dosens of millisconds can be accomplished by formation
of recurrent neural networks. 


We also suggest the solution to the high firing frequency problem by using 
a more precise model of the spike generation mechanism, i.e. the leaky integrator model
with varying in time resting potential and integration time.
The minimum firing frequency for the recurrent network based on such model 
is determined by the rate of variation of the potential and time-constant and
is within the range of physiologically observed values.
Further experimental work is needed to see if the model is applicable
to other types of neurons, such as inhibitory cells.

In conclusion we have studied the stability of delayed activity 
in the recurrent neural network subjected to the influence of noise. 
We conclude that the global stability of the persistent activity 
is affected by properties of synaptic receptor channel. 
NMDA channel, having a long EPSC duration time, is a reliable
mediator of the delayed responce. On the other hand AMPA
receptor is much less reliable, and for the case of synchronized firing
in principle cannot be used to sustain responce. 
Effect of the NMDA channel blockade on the WM task performance is discussed.


\section{ACKNOLEDGEMENTS}

The author is grateful to Thomas Albright, Paul Tiesinga, and Tony Zador
for discussions and numerous helpful suggestions.
This work was supported by the Alfred P. Sloan Foundation.


\appendix
\section{Single-neuron transduction function}
\label{AppendixA}

In this Appendix we calculate the single-neuron transduction function. The 
first step is to find the membrane voltage as a function of time. 
From Eq.~(\ref{SZequation}) using the variation of integration constant we obtain
\begin{equation}
\begin{array}{l}
{\displaystyle V(t) = V_{reset}e^{-S(t)} } \\ \\
{\displaystyle + \int_{t_0}^t dt' e^{S(t')-S(t)} \left[ I(t') + E(t')/\tau(t') \right].}
\end{array}
\end{equation}
Here
\begin{equation}
S(t) = \int_{t_0}^t dt'/\tau (t'),
\label{St}
\end{equation}
and we introduced the refractory period both for generality and to resolve the 
peculiarity at $t=0$.
For constant current and functions given by (\ref{SZE}) and (\ref{SZtau})
the expression for voltage can be further simplified:
\begin{equation}
\begin{array}{l}
{\displaystyle V(t) = V_{reset}e^{-S(t)} + E_0\left[1-e^{-S(t)}\right] } \\ \\ 
{\displaystyle+ \left(I-\frac{\Delta E}{\tau_0}\right)\frac {\tau_0}{\alpha}
\frac{L_{1/\alpha}\left( e^{\alpha t/\tau_0}\right)e^{-S(t)}}
{\left(e^{\alpha t_0/\tau_0} - 1\right)^{1/\alpha}}},
\end{array}
\end{equation}
where
\begin{equation}
S(t) = \frac{1}{\alpha}\ln \left( \frac{e^{\alpha t/\tau_0} - 1}{e^{\alpha t_0/\tau_0} - 1}\right)
\end{equation}
and $L_n(x)$ is defined for integer $n$ by
\begin{equation}
\begin{array}{l}
{\displaystyle L_0(x) = \ln(x),} \\ \\
{\displaystyle L_{n}(x) = \frac{1}{n}(x-1)^n - L_{n-1}(x),}
\end{array}
\end{equation}
and is obtained for fractional $n$ by analytical continuation.
For example
\begin{equation}
\begin{array}{l}
{\displaystyle L_1(x) = x-1-\ln(x),} \\ \\
{\displaystyle L_2(x) = (x-1)^2/2-x+1-\ln(x),} \\ \\
{\displaystyle L_3(x) = (x-1)^3/3-(x-1)^2/2+x-1-\ln(x),} \\ \\ 
{\cdots}
\end{array}
\end{equation}
Solving the equation $V(t)=\theta$ produces the interspike interval $t$
and frequency $f=1/t$. The solution cannot be done in the closed form.
Some asymptotes can be calculated however.
The calculation depends on the value of parameter $\xi = \alpha t/\tau_0$.

{\it i)} $\xi = \alpha t/\tau_0 \ll 1$.

In this limit
\begin{equation}
e^{-S(t)} \approx \frac {\left(e^{\alpha t_0/\tau_0} - 1\right)^{1/\alpha}}{\xi+\xi^2/2},
\end{equation}
is very small and can be neglected everywhere, except for when multiplied
by large factor $L_{1/\alpha}$. The latter product
\begin{equation}
\frac{L_{1/\alpha}\left( e^{\alpha t/\tau_0}\right)e^{-S(t)}}
{\left(e^{\alpha t_0/\tau_0} - 1\right)^{1/\alpha}}
\approx \frac{\alpha}{1+\alpha}\left(\xi+\frac{\xi^2}{2}\right)-\frac{\alpha}{1+2\alpha}\xi^2.
\end{equation}
The equation on the interspike interval is
\begin{equation}
\theta = E_0 +(\tau_0 I - \Delta E) 
\left[ \frac{\xi}{1+\alpha} - \frac{\xi^2}{2} \frac{1}{(1+\alpha)(1+2\alpha)} \right].
\end{equation}
Solving this quadratic equation with respect to $\xi$ we obtain the asymptote given by
Eq.~(\ref{large_frequencies}).

{\it ii)} $\xi \gg 1$.

In this case the solution can be found directly from Eq.~(\ref{SZequation})
by assuming $\dot{V}=0$. Solving the resulting algebraic equation
for $t$ we obtain Eq.~(\ref{small_frequencies}) above.


\section{Decay of the high frequency attractor}
\label{Fluctuations}

\subsection{Asynchronous case: derivation of $N^*_1$ and $N^*_2$}  
                                                
The average and standard deviation of the input current
of each neuron, given by Eqs.~(\ref{totalEPSC}) and (\ref{EPSC})
are
\begin{equation}
\bar{I} = N \kappa f \tau_{_{\rm EPSC}} j_0
\end{equation}
and                                             
\begin{equation}
\overline{\delta I_n^2} = \overline {(I_n-\bar{I})^2} = N \kappa f \tau_{_{\rm EPSC}} j_0^2/2.
\label{deltaIn}
\end{equation}
The distribution of $I_n$ is therefore, according to the central limit theorem
\begin{equation}
p(I_n) = \frac {1}{\sqrt{2\pi\overline{\delta I_n^2}}} 
\exp \left[ -\frac{(I_n - \bar{I})^2}{2\overline{\delta I_n^2}}\right].
\label{InDistribution}
\end{equation}                                             
The probability that the input current is below threshold $I^*$, i.e. the
neuron does not fire, is given by the error function derived from distribution
(\ref{InDistribution})
\begin{equation}
P = \int_{-\infty}^{I^*}dI_n p(I_n) \approx A
e^{-\kappa f \tau_{_{\rm EPSC}} N \Delta I^2/\bar{I}^2}, 
\label{SingleNeuronProbability}
\end{equation}                                     
where $A = \bar{I}/\Delta I\sqrt{4\pi \kappa f \tau_{_{\rm EPSC}}N }$,
and $\Delta I = \bar{I} - I^*$. In derivation of (\ref{SingleNeuronProbability})
we used the asymptotic expression for the error function
\begin{equation}
\int_{-\infty}^{y} dx \frac{e^{-x^2/2\sigma^2}}{\sigma\sqrt{2\pi}} \approx 
\frac {\sigma}{ |y|\sqrt{2\pi}} e^{-y^2/2\sigma^2},
\end{equation}
when $y \ll -\sigma$.

The reverberations of current will be impossible if $n = N\Delta I/\bar{I}$ neurons are below 
threshold. When this occurs the feedback current to each neuron is reduced with respect to the average current by 
$\Delta I$, i.e. is below threshold, and further delayed activity is impossible. 
The probability of this event is given by the binomial distribution:
\begin{equation}
p_0 \sim C_N^n P^n.
\label{p_0}
\end{equation}
Here the binomial coefficient $C_N^n=N!/(N-n)!n!$ accounts for
the large number of groups of neurons that can contribute to the decay.

Since the input currents stay approximately constant during time interval $\tau_{_{\rm EPSC}}$,
we brake the time axis into windows with the duration $\Delta t \sim \tau_{_{\rm EPSC}}$.
Denote by $p$ the average probability of decay of the high frequency activity
during such a little window.
Assume that $k$ windows have been passed since the delayed activity commenced.
The probability that the decay occurs during the $k$-th time window,
i.e. exactly between $t=k\Delta t$ and $t=(k+1)\Delta t$, is
\begin{equation}
\rho(t) \Delta t = (1-p)^k p.
\label{Poisson1}
\end{equation}
Here $(1-p)^k$ is the probability that the decay did not happen during either of $k$ early time
intervals.
After simple manipulations with this expression we obtain
for the density of probability of decay as a function of time 
\begin{equation}
\rho(t) = e^{-p t/\Delta t} / p\Delta t.
\label{Poisson2}
\end{equation}
This is a Poisson distribution and decay is a Poisson process. 
The reason for this is that the system retains the values of input currents
during the time interval $\Delta t \sim \tau_{_{\rm EPSC}}\sim 100$ msec. Therefore all 
processes separated by longer times are independent. Since the 
decay time is of the order of $10$-$100$ seconds, the attempts of system to decay
at various times can be considered independent.
We finally notice that since the values of current are preserved  on the
scales of $\tau_{_{\rm EPSC}}$ we can conclude that
\begin{equation}
p \sim p_0
\end{equation}
given by Eq.~(\ref{p_0}). The average decay time can then be estimated using (\ref{Poisson2})
\begin{equation}
\frac {\bar{t}}{\tau_{_{\rm EPSC}}}\sim p_0^{-1}.
\label{AverageTimeEx}
\end{equation} 

In the limit $f\tau_{_{\rm EPSC}}\gg 1$ the minimum size of the network necessary
to maintain the delayed activity is small. We can assume therefore $N$ to be small
in this limit. The same assumption can be made for $n<N$. Thus for an effective network
(using the neurons sparingly) we can assume the combinatorial term in Eq.~(\ref{p_0}) to be close to unity. 
Since in the expression for $p_0 \sim C_N^n A^n \exp \left(- n \kappa f \tau_{_{\rm EPSC}} N \Delta I^2/\bar{I}^2\right)$
it is also compensated by the small prefactor of the exponential $A^n<1$ [see Eq.~(\ref{SingleNeuronProbability})],
we can assume $C_N^n A^n\sim 1$ for the effective network. 
Therefore in the limit $f\tau_{_{\rm EPSC}}\gg 1$
the main dependence of the parameters of the model is concentrated in the exponential factor.
This set of assumptions in combination with Eq.~(\ref{AverageTimeEx}) leads to the expression (\ref{AverageTime})
in the main text. If more precision is needed Eq.~(\ref{AverageTimeEx}) can be used directly.

To find the critical number of neurons $N^*$ we calculate logarithm of  
both sides of Eq.~(\ref{AverageTimeEx}). We obtain
\begin{equation}
\ln \left( \bar{t}/\tau_{_{\rm EPSC}} \right) \approx -\ln C_{N^*}^n - n\ln p_0.
\end{equation}
Using asymptotic expression for the logarithm of the binomial coefficient, 
which can be derived from the Stirling's formula,
\begin{equation}
\begin{array}{l}
{\displaystyle \ln C_N^n \approx N \phi \left( n/N \right); N, n \gg 1;} \\ \\
{\displaystyle \phi(x) = x\ln \left[1/x\right] + (1-x) \ln \left[1/(1-x)\right]},
\end{array}
\end{equation}
we derive the quadratic equation for $N^*$
\begin{equation}
\begin{array}{l}
{\displaystyle \left( \frac {\Delta I}{I} \right)^3 \kappa f\tau_{_{\rm EPSC}} \left( N^*\right)^2 - } \\ \\
{\displaystyle - N^* \left[ \phi \left( \frac {\Delta I}{I} \right) + \frac {\Delta I}{I} A \right]
- \ln \frac {\bar{t}}{\tau_{_{\rm EPSC}}} \approx 0. }
\end{array}
\end{equation}
Solving this equation we obtain the set of equations (\ref{Nmin1}), (\ref{Nmin}), and (\ref{Nmin2})
in the main text.

\subsection{Synchronous firing case}

In this subsection we assume that all the neurons in the 
network fire simultaneously. The important quantities which will
be studied are the time dependence of the averaged over network current $\bar{I}(t)$
and the fluctuations of the input current into single neuron $\overline{\delta j_n^2}$.
The former is responsible for the shift in the distance to the edge of the attraction basin
(Eq.~(\ref{ShiftS})), the latter determines the average decay time, as follows
from the previous subsection.

Assume that the neuron fired at $t=0$. The average number of EPSC's
arriving to the postsynaptic terminal is $\kappa N$ due to finitness
probability of the synaptic vesicle release $\kappa$ and all-to-all
topology of the connections. The average current at times $0<t<T=1/f$ is 
contributed by the spikes at $t=0$ and by all previous spikes at times $Tk$, $k=1,2,\ldots$:
\begin{equation}
\begin{array}{l}
{\displaystyle \bar{I}(t) = I_0e^{-t/\tau_{_{\rm EPSC}}}+I_0e^{-(t+T)/\tau_{_{\rm EPSC}}} + \ldots } \\ \\
{\displaystyle  = I_0e^{-t/\tau_{_{\rm EPSC}}}/(1-e^{-1/f\tau_{_{\rm EPSC}}})},
\end{array}
\label{GeometricI}
\end{equation}
where we introduced the notation $I_0 = \kappa N j_0$.
The average over time value of this averaged over neurons current is 
$<\bar{I}>=I_0f\tau_{_{\rm EPSC}}$,
where by angular brackets we denote the time average: $<A> = \int_0^T A(t) dt/T$.
The average current $\bar{I}$ therefore experiences oscillations
between $\bar{I}_{max} = \bar{I}(0)$ and $\bar{I}_{min} = \bar{I}(T)$.
The minimum value of $\bar{I}$ is below the average current $<\bar{I}>$.
This minimum value, according to the logic presented in the main text, determines 
the shift of the distance to the edge of the attraction basin. This
shift is therefore $<\bar{I}> - \bar{I}_{min}$. Our computer modeling
however shows that the actual amplitude of the current oscillations is
consistently about $60$\% of the value predicted by this argument.
This was tested at various values of parameters.
The explanation of this $60$\% factor is as follows. 
In case if all the neurons fire simultaneously
the shape of the dependence of the average current versus time 
is saw-like. However the spikes do not fire absolutely
simultaneously. The uncertainty in the spiking time is 
of the order of $T$. This uncertainty, which is intrinsic to
the recurrent synaptic noises, and therefore is difficult to calculate
exactly, smears out the saw-like dependence of the average current
of time (see inset in Figure~\ref{fig25}a). Approximately this smearing can be accounted for by
dumping the higher harmonics of the saw-like dependence. When only
principal harmonic remains, the amplitude of the saw-like curve
is reduced by a factor $2/pi\approx 0.6$. This is consistent
with the numerical result. We therefore conclude that the shift
of $\Delta I$ is 
\begin{equation}
\delta I_{syn} \approx 0.6 \left[ <\bar{I}> - \bar{I}(T)\right],
\end{equation}
what leads to Eq.~(\ref{deltaIs}) in the text. 

We now turn to the calculation of the standard deviation of the 
input current into one neuron. 
Similar to (\ref{GeometricI}), using the central limit theorem
we obtain
\begin{equation}
\delta I_n^2(t) = \sum_{k=0}^{\infty} j_0^2\kappa N e^{-(2t+2Tk)/\tau_{_{\rm EPSC}}}
\end{equation}
This quantity is most important at $t \approx T$ when the average current reaches its minimum and
the neuron stops firing with maximum probability.
Performing the summation we therefore obtain 
\begin{equation}
\delta I_n^2(T) = \frac {\kappa I^2}{(\kappa f \tau_{_{\rm EPSC}} )^2 N} 
\frac {1} {e^{2/f\tau_{_{\rm EPSC}}}-1}.
\end{equation}
Substituting this value into Eq.~(\ref{SingleNeuronProbability}) in the
previous subsection and repeating the subsequent derivation 
we obtain Eqs.~(\ref{Nmin1s}) and (\ref{Nmin2s}) in the main text.

\subsection{Fluctuations of the average current}                                               

In this subsection we calculate the fluctuations of the average 
netwotk current. We consider asynchronous case for simplicity.
The conclusions are perfectly good for synchronous case, 
for the reasons that will become clear later in this subsection, 
and agree well with computer similations.

The average current satisfies linarized equation similar to linearized version
of Eq.~(\ref{MFequation})
\begin{equation}
\tau_{_{\rm EPSC}} \Delta \dot{\bar{I}}(t) = (\nu - 1) \Delta \bar{I}(t) + \xi(t).
\label{LinI}
\end{equation}
Here $\Delta \bar{I}(t)$ is the deviation of the averaged over network current
from the equilibrium value and $\xi(t)$ is the noise. The unitless 
network feedback coefficient $\nu < 1$ is defined by (\ref{FeedbackCoefficient}).
As evident
from this equation $\Delta \bar{I}(t)$ has a slow time constant $\tau_{_{\rm EPSC}}/(1-\nu)$
[since $1/(1-\nu)>5$ in our simulations]. On the other hand the noise $\xi(t)$
is determined by synapses and has a correlation time that is relatively small 
($\sim \tau_{_{\rm EPSC}}$). 

The correlation function of noise in the average current can be found 
from Eq.~(\ref{totalEPSC})
\begin{equation}
\Xi(t)\equiv \left<\xi(t)\xi(0)\right> = \frac{\overline{\delta I_n^2}}{N} \exp(-|t|/\tau_{_{\rm EPSC}}).
\end{equation}
Here angular brackets imply averaging over time and the value
of $\Xi(t=0)$ follows from Eq.~(\ref{deltaIn}) and the central
limit theorem (dispersion of the average is equal to the dispersion of each of the 
homogenious constituents $I_n$ divided by the number of elements $N$).
We conclude therefore that
\begin{equation}
\Xi(t=0) = \frac{I^2}{2f\tau_{_{\rm EPSC}}N^2}.
\end{equation}
The correlation function of $\Delta \bar{I}$ can then be easily found 
from Eq.~(\ref{LinI}) using the Fourier transform.
Define $C(t)\equiv \left< \Delta \bar{I}(t) \Delta \bar{I}(0) \right>$.
Then 
\begin{equation}
C(\omega) = \left<\left| \Delta \bar{I} (\omega) \right|^2\right>
= \frac{\Xi(\omega)}{(1-\nu)^2 + \left(\omega \tau_{_{\rm EPSC}}\right)^2}
\label{Comega}
\end{equation} 
Since 
\begin{equation}
\Xi(\omega) = -\frac{2i\tau_{_{\rm EPSC}}\Xi(t=0)}{1+\left(\omega \tau_{_{\rm EPSC}}\right)^2},
\end{equation}
the expression for $C(t)$ is readily obtained by inverting the Fourier 
transform (\ref{Comega}).
In the limit $1-\nu \ll 1$, which holds in our simulations the answer is
\begin{equation}
C(t) = \frac{\Xi(t=0)}{1-\nu} \exp\left[-(1-\nu) |t|/\tau_{_{\rm EPSC}} \right].
\label{answC}
\end{equation}
The value of $C$ taken at $t=0$ determines the standard deviation of the 
average current (\ref{deltaI}).

The fluctuations described by the correlation function (\ref{answC})
have a large correlation time compared to the firing frequency.
We conclude therefore that synchrony should not affect the long range
component of the correlation function.







\end{multicols}
\end{document}